\documentclass[12pt]{article}
\usepackage{amsmath}

\begin{document}

\baselineskip=17pt

\begin{titlepage}
\rightline{\tt arXiv:0704.0936}
\rightline{\tt DESY 07-047} 
\begin{center}
\vskip 3.0cm
{\Large \bf {Analytic solutions for marginal deformations}}\\
\vskip 0.4cm
{\Large \bf {in open superstring field theory}}
\vskip 1.0cm
{\large {Yuji Okawa}}
\vskip 1.0cm
{\it {DESY Theory Group}}\\
{\it {Notkestrasse 85}}\\
{\it {22607 Hamburg, Germany}}\\
yuji.okawa@desy.de

\vskip 3.0cm

{\bf Abstract}
\end{center}

\noindent
We extend the calculable analytic approach
to marginal deformations
recently developed in open bosonic string field theory
to open superstring field theory formulated by Berkovits.
We construct analytic solutions to all orders
in the deformation parameter
when operator products
made of the marginal operator
and the associated superconformal primary field
are regular.

\medskip

\end{titlepage}

\newpage


\section{Introduction}
\setcounter{equation}{0}

Ever since the analytic solution for tachyon condensation
in open bosonic string field theory~\cite{Witten:1985cc}
was constructed by Schnabl \cite{Schnabl:2005gv},
new analytic technologies have been developed
\cite{Okawa:2006vm, Fuchs:2006hw, Fuchs:2006an, Rastelli:2006ap,
Ellwood:2006ba, Fuji:2006me, Fuchs:2006gs, Okawa:2006sn,
Asano:2006hk, Asano:2006hm, Erler:2006hw, Imbimbo:2006tz,
Erler:2006ww},
and analytic solutions for marginal deformations
were recently constructed \cite{Schnabl:2007az,
Kiermaier:2007ba}.\footnote{
For earlier study of marginal deformations in string field theory
and related work,
see \cite{Sen:2000hx, Iqbal:2000qg, Takahashi:2001pp, Marino:2001ny,
Kluson:2002ex, Takahashi:2002ez, Kluson:2002hr, Kluson:2002av,
Kluson:2003xu, Coletti:2003ai, Berkovits:2003ny, Sen:2004cq,
Katsumata:2004cc, Yang:2005iu, Kishimoto:2005bs}.
}
We believe that we are now in a new phase of research
on open string field theory.\footnote{
See \cite{Taylor:2003gn, Sen:2004nf, Rastelli:2005mz,
Taylor:2006ye} for reviews.}

Extension of these new technologies
to closed string field theory, however,
does not seem straightforward.
The star product \cite{Witten:1985cc}
used in open string field theory
has a simpler description
in the conformal field theory (CFT) formulation
when we use a coordinate called the sliver frame
which was originally introduced in \cite{Rastelli:2000iu}.
It has been an important ingredient
in recent developments.
Closed bosonic string field theory \cite{Zwiebach:1992ie,
Saadi:1989tb, Kugo:1989aa, Kugo:1989tk, Kaku:1988zv, Kaku:1988zw}
and heterotic string field theory
\cite{Okawa:2004ii, Berkovits:2004xh}, however, use
infinitely many non-associative string products,
and we have not found any coordinate
where simple descriptions of these string products
are possible.

On the other hand, extension to open superstring field theory
formulated by Berkovits \cite{Berkovits:1995ab} is promising
because the string product used in the theory
is the same as that in open bosonic string field theory.
In this paper we construct analytic solutions
for marginal deformations
in open superstring field theory.

We first review
the solutions for marginal deformations
in open bosonic string field theory.
The solutions take the form of an expansion
in terms of the deformation parameter $\lambda$,
and analytic expressions to all order in $\lambda$
have been derived
when operator products made of the marginal operator
are regular \cite{Schnabl:2007az, Kiermaier:2007ba}.
When the operator product of the marginal operator with itself
is singular,
solutions were constructed to $O(\lambda^3)$
by regularizing the singularity
and by adding counterterms \cite{Kiermaier:2007ba}.

The goal of this paper is to construct
analytic solutions in open superstring field theory
when operator products
made of the marginal operator
and the associated superconformal primary field
of dimension $1/2$ are regular.
It will be a starting point
for constructing analytic solutions
when these operators have singular operator products.
We first simplify the equation of motion
for open superstring field theory by field redefinition.
We then make an ansatz
motivated by the structure of the solutions
in the bosonic case
and solve the equation of motion analytically.
The solutions in the superstring case
turn out to be remarkably simple
and similar to those in the bosonic case.
The final section of the paper is devoted
to conclusions and discussion.

\medskip

We learned that T.~Erler independently
found analytic solutions for marginal deformations
in open superstring field theory \cite{Erler:2007rh}
prior to our construction.

\section{Solutions in open bosonic string field theory}
\setcounter{equation}{0}

In this section, we review the analytic solutions
for marginal deformations
constructed in \cite{Schnabl:2007az, Kiermaier:2007ba}
for the open bosonic string.
The equation of motion for open bosonic string field theory
\cite{Witten:1985cc} is given by
\begin{equation}
Q_B \Psi + \Psi^2 = 0 \,,
\label{bosonic-equation-of-motion}
\end{equation}
where $\Psi$ is the open string field
and $Q_B$ is the BRST operator.
All the string products in this paper
are defined by the star product \cite{Witten:1985cc}.
The open bosonic string field $\Psi$ has
ghost number $1$ and is Grassmann odd.
The BRST operator is Grassmann odd
and is nilpotent: $Q_B^2 = 0 \,$.
It is a derivation with respect to the star product:
\begin{equation}
Q_B \, ( \varphi_1 \, \varphi_2 )
= (Q_B \varphi_1) \, \varphi_2
+ (-1)^{\varphi_1} \, \varphi_1 \, (Q_B \, \varphi_2)
\end{equation}
for any states $\varphi_1$ and $\varphi_2$,
where $(-1)^{\varphi_1} = 1$ when $\varphi_1$ is Grassmann even
and $(-1)^{\varphi_1} = -1$ when $\varphi_1$ is Grassmann odd.

The deformation of the boundary CFT for the open string
by a matter primary field $V$ of dimension $1$ is marginal
to linear order in the deformation parameter.
When the deformation is exactly marginal,
we expect a solution of the form
\begin{equation}
\Psi_\lambda = \sum_{n=1}^\infty \lambda^n \, \Psi^{(n)} \,,
\end{equation}
where $\lambda$ is the deformation parameter,
to the nonlinear equation of motion
(\ref{bosonic-equation-of-motion}).
When operator products made of $V$ are regular,
analytic expressions of $\Psi^{(n)}$'s were
derived in \cite{Schnabl:2007az, Kiermaier:2007ba},
and the BPZ inner product
$\langle \, \varphi, \Psi^{(n)} \, \rangle$
for a state $\varphi$ in the Fock space is given by
\begin{equation}
\begin{split}
\langle \, \varphi, \Psi^{(n)} \, \rangle
& = \int_0^1 dt_1 \int_0^1 dt_2 \ldots \int_0^1 dt_{n-1} \,
\langle \, f \circ \varphi (0) \, c V (1) \, {\cal B} \,
c V (1+t_1) \, {\cal B} \, c V (1+t_1+t_2) \, \ldots \\
& \qquad \qquad \qquad \qquad \qquad \qquad \qquad {}\times
{\cal B} \, c V (1+t_1+t_2+ \ldots +t_{n-1}) \,
\rangle_{{\cal W}_{1+t_1+t_2+ \ldots +t_{n-1}}} \,.
\end{split}
\end{equation}
We follow the notation used in \cite{Okawa:2006vm, Okawa:2006sn,
Kiermaier:2007ba}. In particular, see the beginning of section 2
of \cite{Okawa:2006vm} for the relation to the notation
used in \cite{Schnabl:2005gv}. 
Here and in what follows we use $\varphi$ to denote
a generic state in the Fock space
and $\varphi (0)$ to denote its corresponding operator
in the state-operator mapping.
We use the doubling trick
in calculating CFT correlation functions.
As in \cite{Okawa:2006sn},
we define the oriented straight lines $V^\pm_\alpha$ by
\begin{equation}
\begin{split}
& V^\pm_\alpha = \Bigl\{ \, z \, \Big| \,
{\rm Re} (z) = {}\pm \frac{1}{2} \, (1+\alpha) \, \Bigr\} \,, \\
& \hbox{orientation} \, :
{}\pm \frac{1}{2} \, (1+\alpha) -i \, \infty
\to {}\pm \frac{1}{2} \, (1+\alpha) +i \, \infty \,,
\end{split}
\end{equation}
and the surface ${\cal W}_\alpha$ can be represented
as the region between $V^-_0$ and $V^+_{2 \alpha}$,
where $V^-_0$ and $V^+_{2 \alpha}$ 
are identified by translation.
The function $f(z)$ is
\begin{equation}
f(z) = \frac{2}{\pi} \, \arctan \, z \,,
\end{equation}
and $f \circ \varphi (z)$ denotes
the conformal transformation of $\varphi (z)$ by the map $f(z)$.
The operator ${\cal B}$ is defined by
\begin{equation}
{\cal B} = \int \frac{dz}{2 \pi i} \, b(z) \,,
\end{equation}
and when ${\cal B}$ is located between two operators
at $t_1$ and $t_2$ with $1/2 < t_1 < t_2$,
the contour of the integral can be taken to be $-V^+_\alpha$
with $2 \, t_1 - 1 < \alpha < 2 \, t_2 - 1$.
The anticommutation relation of ${\cal B}$ and $c(z)$ is
\begin{equation}
\{ {\cal B}, c(z) \} = 1 \,,
\end{equation}
and ${\cal B}^2 = 0$.

The solution can be written more compactly as
\begin{equation}
\langle \, \varphi, \Psi^{(n)} \, \rangle
= \int_0^1 dt_1 \int_0^1 dt_2 \ldots \int_0^1 dt_{n-1} \,
\Bigl\langle \, f \circ \varphi (0) \,
\prod_{i=0}^{n-2} \Bigl[ \, c V (1+ \ell_i) \, {\cal B} \,
\Bigr] \, c V (1+ \ell_{n-1}) \,
\Bigr\rangle_{{\cal W}_{1+\ell_{n-1}}} \,,
\end{equation}
where
\begin{equation}
\ell_0 = 0 \,, \quad \ell_i \equiv \sum_{k=1}^i t_k \quad
\mbox{for} \quad i=1 \,, \, 2 \,, \, 3 \,, \, \ldots \,.
\label{ell-definition}
\end{equation}
It can be further simplified as
\begin{equation}
\Psi_\lambda
= \frac{1}{1- \lambda \, X_b \, J_b} \, \lambda \, X_b \,,
\label{bosonic-solution}
\end{equation}
where
\begin{equation}
\frac{1}{1- \lambda \, X_b \, J_b}
\equiv 1 + \sum_{n=1}^\infty \,
( \, \lambda \, X_b \, J_b \, )^n \,.
\end{equation}
The state $X_b$ is the same as $\Psi^{(1)}$:
\begin{equation}
\langle \, \varphi, X_b \, \rangle
= \langle \, f \circ \varphi (0) \, c V (1) \,
\rangle_{{\cal W}_1} \,.
\end{equation}
It solves the linearized equation of motion: $Q_B X_b = 0$.
The definition of $J_b$ is a little involved.
It is defined when it appears as $\varphi_1 \, J_b \, \varphi_2$
between two states $\varphi_1$ and $\varphi_2$ in the Fock space.
The string product $\varphi_1 \, J_b \, \varphi_2$ is given by
\begin{equation}
\langle \, \varphi,\, \varphi_1 \, J_b \,\varphi_2 \, \rangle
= \int_0^1 dt \, \langle \, f \circ \varphi (0) \,
f_1 \circ \varphi_1 (0) \, {\cal B} \, f_{1+t} \circ \varphi_2 (0) \,
\rangle_{{\cal W}_{1+t}} \,,
\end{equation}
where $\varphi_1 (0)$ and $\varphi_2 (0)$
are the operators
corresponding to the states $\varphi_1$ and $\varphi_2$, respectively.
The map $f_a (z)$ is a combination of $f(z)$ and translation:
\begin{equation}
f_a (z) = \frac{2}{\pi} \, \arctan \, z + a \,.
\end{equation}
The string product $\varphi_1 \, J_b \, \varphi_2$
is well defined if
$f_1 \circ \varphi_1 (0) \, {\cal B} \, f_{1+t} \circ \varphi_2 (0)$
is regular in the limit $t \to 0 \,$.
In the definition of $\Psi_\lambda$,
$J_b$ always appears between two $X_b$'s.
Since
$c(1) \, {\cal B} \, c(1+t) = c(1)$ in the limit $t \to 0 \,$,
the ghost part of $X_b \, J_b \, X_b$ is finite.\footnote{
Note that $f_a \circ cV(0) = cV(a)$
because $cV$ is a primary field of dimension $0$.
}
Therefore, $X_b \, J_b \, X_b$ is well defined
if the operator product $V(1) \, V(1+t)$ is regular
in the limit $t \to 0 \,$.
The ghost part of the state $\Psi^{(n)} = ( X_b \, J_b )^{n-1} X_b$
is also finite
because ${\cal B} \, c(z) \, {\cal B} = {\cal B}$
and $c(1) \, {\cal B} \, c(1+\ell_{n-1}) = c(1)$
in the limit $\ell_{n-1} \to 0 \,$.
Therefore, $\Psi^{(n)}$ is well defined if the operator product
in the matter sector
\begin{equation}
\int_0^1 dt_1 \int_0^1 dt_2 \ldots \int_0^1 dt_{n-1} \,
\prod_{i=0}^{n-1} \Bigl[ \, V (1+ \ell_i) \, \Bigr]
\end{equation}
is finite.
For example, the marginal deformation
associated with the rolling tachyon
and the deformations in the light-cone directions
satisfy the regularity condition
\cite{Schnabl:2007az, Kiermaier:2007ba}.

An important property of $J_b$ is
\begin{equation}
\varphi_1 \, (Q_B J_b) \, \varphi_2
= \varphi_1 \, \varphi_2
\label{Q_B-J_b}
\end{equation}
when $f_1 \circ \varphi_1 (0) \, f_{1+t} \circ \varphi_2 (0)$
vanishes in the limit $t \to 0 \,$.
Since the BRST transformation of $b (z)$
is the energy-momentum tensor $T(z)$,
the inner product $\langle \, \varphi,\,
\varphi_1 \, (Q_B J_b) \,\varphi_2 \, \rangle$
is given by
\begin{equation}
\langle \, \varphi,\, \varphi_1 \, (Q_B J_b) \,\varphi_2 \, \rangle
= \int_0^1 dt \, \langle \, f \circ \varphi (0) \,
f_1 \circ \varphi_1 (0) \, {\cal L} \, f_{1+t} \circ \varphi_2 (0) \,
\rangle_{{\cal W}_{1+t}} \,,
\end{equation}
where
\begin{equation}
{\cal L} = \int \frac{dz}{2 \pi i} \, T(z) \,,
\end{equation}
and the contour of the integral is the same as that of ${\cal B}$.
As discussed in \cite{Okawa:2006vm},
an insertion of ${\cal L}$ is equivalent
to taking a derivative with respect to $t$.
It is analogous to the relation
$L_0 \, e^{-t L_0} = {}-\partial_t \, e^{-t L_0}$
in the standard strip coordinates,
where $L_0$ is the zero mode of the energy-momentum tensor.
We thus have
\begin{equation}
\begin{split}
\langle \, \varphi,\, \varphi_1 \, (Q_B J_b) \,\varphi_2 \, \rangle
& = \int_0^1 dt \, \partial_t \, \langle \, f \circ \varphi (0) \,
f_1 \circ \varphi_1 (0) \, f_{1+t} \circ \varphi_2 (0) \,
\rangle_{{\cal W}_{1+t}} \\
& = \langle \, f \circ \varphi (0) \,
f_1 \circ \varphi_1 (0) \, f_2 \circ \varphi_2 (0) \,
\rangle_{{\cal W}_2}
\end{split}
\end{equation}
when $f_1 \circ \varphi_1 (0) \, f_{1+t} \circ \varphi_2 (0)$
vanishes in the limit $t \to 0 \,$.
This completes the proof of ({\ref{Q_B-J_b}).
When $\varphi_1 = \varphi_2 = X_b$,
the operator product $c V (1) \, c V (1+t)$ vanishes
in the limit $t \to 0$
if $V(1) \, V(1+t)$ is regular in the limit $t \to 0 \,$.
In the language of $\cite{Kiermaier:2007ba}$,
$\varphi_1 \, J_b \, \varphi_2$ is
\begin{equation}
\varphi_1 \, J_b \, \varphi_2
= \int_0^1 dt \, \varphi_1 \, e^{-(t-1) L^+_L} \,
(-B^+_L) \, \varphi_2 \,,
\end{equation}
and the relation (\ref{Q_B-J_b}) follows from
$\{ Q_B ,\, B^+_L \} = L^+_L$.

To summarize, when operator products made of $V$ are regular,
the solution (\ref{bosonic-solution}) is well defined,
and we can safely use the relations
\begin{equation}
Q_B X_b = 0 \,, \qquad Q_B J_b = 1
\end{equation}
for the Grassmann-odd states $X_b$ and $J_b$
when we calculate the BRST transformation of $\Psi_\lambda$.
It is now straightforward to calculate $Q_B \Psi_\lambda$,
and the result is
\begin{equation}
Q_B \Psi_\lambda
= {}- \frac{1}{1- \lambda \, X_b \, J_b} \, \lambda \, X_b \,
\frac{1}{1- \lambda \, X_b \, J_b} \, \lambda \, X_b \,.
\end{equation}
We have thus shown that $\Psi_\lambda$ in (\ref{bosonic-solution})
satisfies the equation of motion
(\ref{bosonic-equation-of-motion}).

\section{Equation of motion for open superstring field theory}
\setcounter{equation}{0}

The equation of motion
for open superstring field theory \cite{Berkovits:1995ab} is
\begin{equation}
\eta_0 \, ( \, e^{-\Phi} \, Q_B \, e^\Phi \, ) = 0 \,,
\end{equation}
where $\Phi$ is the open superstring field.
It is Grassmann even
and has ghost number $0$ and picture number $0$.
The superghost sector is described by
$\eta$, $\xi$, and $\phi$ \cite{Friedan:1985ge, Polchinski:1998rr},
and the zero modes of $\eta$ and $\xi$ are included
in the Hilbert space.
The operator $\eta_0$ is the zero mode of $\eta$
and a derivation with respect to the star product.
For any states $\varphi_1$ and $\varphi_2$, we have
\begin{equation}
\eta_0 \, ( \varphi_1 \, \varphi_2 )
= (\eta_0 \, \varphi_1) \, \varphi_2
+ (-1)^{\varphi_1} \, \varphi_1 \, (\eta_0 \, \varphi_2) \,,
\end{equation}
as in the case of $Q_B$,
where $(-1)^{\varphi_1} = 1$ when $\varphi_1$ is Grassmann even
and $(-1)^{\varphi_1} = -1$ when $\varphi_1$ is Grassmann odd.
The Grassmann-odd operator $\eta_0$ is nilpotent
and anticommutes with $Q_B$:
\begin{equation}
Q_B^2 = 0 \,, \qquad
\eta_0^2 = 0 \,, \qquad
\{ Q_B ,\, \eta_0 \} = 0 \,.
\end{equation}

Since $\eta_0 \, ( \, e^{-\Phi} \, Q_B \, e^\Phi \, )
= e^{-\Phi} \, [ \,
Q_B \, ( \, e^\Phi \, \eta_0 \, e^{-\Phi} \, ) \, ] \, e^\Phi \,$,
the equation of motion can also be written as follows:
\begin{equation}
Q_B \, ( \, e^\Phi \, \eta_0 \, e^{-\Phi} \, ) = 0 \,.
\label{old-equation}
\end{equation}
We further simplify the equation of motion by field redefinition.
Since the open superstring field $\Phi$ has
vanishing ghost and picture numbers,
there is a natural class of field redefinitions given by
\begin{equation}
\Phi_{new} = \sum_{n=1}^\infty \, a_n \Phi_{old}^n \,,
\end{equation}
where $a_n$'s are constants.
The map from $\Phi_{old}$ to $\Phi_{new}$
is well defined at least perturbatively.
We choose
\begin{equation}
1 - \Phi_{new} = e^{-\Phi_{old}} \,,
\label{field-redefinition}
\end{equation}
and the equation of motion (\ref{old-equation})
written in terms of $\Phi_{new}$ is
\begin{equation}
{}- Q_B \, \Bigl( \, \frac{1}{1-\Phi} \, \eta_0 \, \Phi \Bigr)
= {}- \frac{1}{1-\Phi} \,
\Bigl[ \, Q_B \, \eta_0 \, \Phi
+ ( \, Q_B \, \Phi \, ) \, \frac{1}{1-\Phi} \,
( \, \eta_0 \, \Phi \, ) \, \Bigr] = 0 \,,
\end{equation}
where
\begin{equation}
\frac{1}{1-\Phi} \equiv 1 + \sum_{n=1}^\infty \, \Phi^n \,.
\end{equation}
In the following sections,
we solve the equation of motion of the form
\begin{equation}
Q_B \, \eta_0 \, \Phi
+ ( \, Q_B \, \Phi \, ) \, \frac{1}{1-\Phi} \,
( \, \eta_0 \, \Phi \, ) = 0 \,,
\label{equation-of-motion}
\end{equation}
or
\begin{equation}
Q_B \, \eta_0 \, \Phi
+ ( \, Q_B \, \Phi \, ) \, ( \, \eta_0 \, \Phi \, )
+ \sum_{n=1}^\infty \, ( \, Q_B \, \Phi \, ) \,
\Phi^n  \, ( \, \eta_0 \, \Phi \, ) = 0 \,.
\end{equation}

\section{Solutions to second order}
\setcounter{equation}{0}

For any marginal deformation of the boundary CFT
for the open superstring,
there is an associated superconformal primary field $V_{1/2}$
of dimension $1/2$,
and the marginal operator $V_1$ of dimension $1$
is the supersymmetry transformation of $V_{1/2}$.
For example, $V_{1/2}$ is the fermionic coordinate $\psi^\mu (z)$
when $V_1$ is the derivative of the bosonic coordinate
$i \, \partial X^\mu (z)$
up to a normalization constant.
In the RNS formalism,
the unintegrated vertex operator in the $-1$ picture
is $c e^{-\phi} V_{1/2}$,
and the unintegrated vertex operator in the $0$ picture
is $c V_1$.
In open superstring field theory \cite{Berkovits:1995ab},
the solution to the linearized equation of motion
$Q_B \, \eta_0 \, \Phi^{(1)} = 0$
associated with the marginal deformation
is given by $\Phi^{(1)} = X$,
where $X$ is the state corresponding
to the operator ${\cal V} (0) = c \, \xi e^{-\phi} V_{1/2} (0)$:
\begin{equation}
\langle \, \varphi, X \, \rangle
= \langle \, f \circ \varphi (0) \,
{\cal V} (1) \, \rangle_{{\cal W}_1}
= \langle \, f \circ \varphi (0) \,
c \, \xi e^{-\phi} V_{1/2} (1) \, \rangle_{{\cal W}_1} \,.
\end{equation}
See \cite{Berkovits:2003ny} for some explicit calculations
in open superstring field theory
when $V_{1/2} (z) = \psi^\mu (z) \,$.

When the deformation is exactly marginal,
we expect a solution of the form
\begin{equation}
\Phi_\lambda = \sum_{n=1}^\infty \lambda^n \, \Phi^{(n)} \,,
\end{equation}
where $\lambda$ is the deformation parameter,
to the nonlinear equation of motion
(\ref{equation-of-motion}).
The equation for $\Phi^{(2)}$ is
\begin{equation}
Q_B \, \eta_0 \, \Phi^{(2)}
= {}- ( Q_B \, \Phi^{(1)} ) \, ( \eta_0 \, \Phi^{(1)} )
= {}- ( Q_B X ) \, ( \eta_0 X ) \,.
\label{Phi^(2)-equation}
\end{equation}
The right-hand side is annihilated by $Q_B$ and by $\eta_0$
because $Q_B \eta_0 X = 0 \,$.
In order to solve the equation for $\Phi^{(2)}$,
we introduce a state $J$
by replacing $b (z)$ in $J_b$ for the bosonic case
with $\xi b (z)$. Since
\begin{equation}
\eta_0 \cdot \xi b (z)
\equiv \oint \frac{dw}{2 \pi i} \, \eta (w) \, \xi b (z)
= b(z)
\label{eta_0-J}
\end{equation}
and the BRST transformation of $b(z)$ gives
the energy-momentum tensor,
we expect that $\xi b (z)$ in the superstring case
plays a similar role of $b (z)$ in the bosonic case.
In fact, the zero mode of $\xi b (z)$ divided by $L_0$
was used in the calculation of on-shell four-point amplitudes
in \cite{Berkovits:1999bs}.
We again define $J$ when it appears
as $\varphi_1 \, J \, \varphi_2$
between two states $\varphi_1$ and $\varphi_2$ in the Fock space.
The string product $\varphi_1 \, J \, \varphi_2$ is given by
\begin{equation}
\langle \, \varphi,\, \varphi_1 \, J \,\varphi_2 \, \rangle
= \int_0^1 dt \, \langle \, f \circ \varphi (0) \,
f_1 \circ \varphi_1 (0) \, {\cal J} \, f_{1+t} \circ \varphi_2 (0) \,
\rangle_{{\cal W}_{1+t}} \,,
\end{equation}
where $\varphi_1 (0)$ and $\varphi_2 (0)$
are the operators corresponding to the states
$\varphi_1$ and $\varphi_2$, respectively.
The operator ${\cal J}$ is defined by
\begin{equation}
{\cal J} = \int \frac{dz}{2 \pi i} \, \xi b(z) \,,
\end{equation}
and when ${\cal J}$ is located between two operators
at $t_1$ and $t_2$ with $1/2 < t_1 < t_2$,
the contour of the integral can be taken to be $-V^+_\alpha$
with $2 \, t_1 - 1 < \alpha < 2 \, t_2 - 1$.
As in the case of $J_b$,
the string product $\varphi_1 \, J \, \varphi_2$
is well defined if
$f_1 \circ \varphi_1 (0) \, {\cal J} \, f_{1+t} \circ \varphi_2 (0)$
is regular in the limit $t \to 0 \,$.
We also have an important relation
\begin{equation}
\varphi_1 \, ( \, Q_B \, \eta_0 \, J \, ) \, \varphi_2
= \varphi_1 \, \varphi_2
\label{Q_B-eta_0-J}
\end{equation}
if $f_1 \circ \varphi_1 (0) \, f_{1+t} \circ \varphi_2 (0)$
vanishes in the limit $t \to 0 \,$.
The proof of this relation follows from that of (\ref{Q_B-J_b})
after we use (\ref{eta_0-J}) in calculating $\eta_0 J \,$.
We will discuss these regularity conditions later
and proceed for the moment assuming they are satisfied.
Namely, we assume that states involving $J$ are well defined
and that we can use the relations
\begin{equation}
Q_B \eta_0 X = 0 \,, \qquad Q_B \eta_0 J = 1
\end{equation}
for the Grassmann-even states $X$ and $J$.

Motivated by the structure of the solutions in the bosonic case,
we look for a solution
which consists of $X \, J \, X$, $Q_B$, and $\eta_0$
to the equation (\ref{Phi^(2)-equation}) for $\Phi^{(2)}$.
There are nine possible states:
\begin{equation}
\begin{array}{lll}
  (Q_B \eta_0 X) \, J \, X = 0 \,, \qquad
& (Q_B X) \, (\eta_0 J) \, X \,, \qquad
& (Q_B X) \, J \, (\eta_0 X) \,, \\
  (\eta_0 X) \, (Q_B J) \, X \,, \qquad
& X \, (Q_B \eta_0 J) \, X = X^2 \,, \qquad
& X \, (Q_B J) \, (\eta_0 X) \,, \\
  (\eta_0 X) \, J \, (Q_B X) \,, \qquad
& X \, (\eta_0 J) \, (Q_B X) \,, \qquad
& X \, J \, (Q_B \eta_0 X) = 0 \,.
\end{array}
\end{equation}
Two of them vanish and one of them reduces to $X^2$.
We then calculate the action of $Q_B \eta_0$
on the nonvanishing states:
\begin{equation}
\begin{split}
Q_B \eta_0 \, [ \, (Q_B X) \, (\eta_0 J) \, X \, ]
& = {}- (Q_B X) \, (\eta_0 X) \,, \\
Q_B \eta_0 \, [ \, (Q_B X) \, J \, (\eta_0 X) \, ]
& = (Q_B X) \, (\eta_0 X) \,, \\
Q_B \eta_0 \, [ \, (\eta_0 X) \, (Q_B J) \, X \, ]
& = {}- (\eta_0 X) \, (Q_B X) \,, \\
Q_B \eta_0 \, [ \, X \, (Q_B \eta_0 J) \, X \, ]
& = {}- (\eta_0 X) \, (Q_B X) + (Q_B X) \, (\eta_0 X) \,, \\
Q_B \eta_0 \, [ \, X \, (Q_B J) \, (\eta_0 X) \, ]
& = {}- (Q_B X) \, (\eta_0 X) \,, \\
Q_B \eta_0 \, [ \, (\eta_0 X) \, J \, (Q_B X) \, ]
& = (\eta_0 X) \, (Q_B X) \,, \\
Q_B \eta_0 \, [ \, X \, (\eta_0 J) \, (Q_B X) \, ]
& = {}- (\eta_0 X) \, (Q_B X) \,.
\end{split}
\end{equation}
We thus find that
$(Q_B X) \, (\eta_0 J) \, X$,
${}- (Q_B X) \, J \, (\eta_0 X)$,
and $X \, (Q_B J) \, (\eta_0 X)$
solve the equation~(\ref{Phi^(2)-equation}) for $\Phi^{(2)}$.
We can also take an appropriate linear combination
of the seven states,
and different solutions should be related
by gauge transformations.
We choose
\begin{equation}
\Phi^{(2)} = ( Q_B X ) \, ( \eta_0 J ) \, X
\label{Phi^(2)}
\end{equation}
and consider its extension to $\Phi^{(n)}$ in the next section.

\section{Solutions in open superstring field theory}
\setcounter{equation}{0}

Remarkably, a simple extension of $\Phi^{(2)}$ in (\ref{Phi^(2)})
solves the equation of motion (\ref{equation-of-motion})
to all orders in $\lambda$.
A solution is given by
\begin{equation}
\begin{split}
\Phi^{(3)} & = ( Q_B X ) \, ( \eta_0 J ) \,
( Q_B X ) \, ( \eta_0 J ) \, X \,, \\
\Phi^{(4)} & = ( Q_B X ) \, ( \eta_0 J ) \,
( Q_B X ) \, ( \eta_0 J ) \, ( Q_B X ) \, ( \eta_0 J ) \, X \,, \\
& \qquad \vdots \\
\Phi^{(n)} & = [ \, ( Q_B X ) \, ( \eta_0 J ) \, ]^{n-1} \, X \,,
\end{split}
\end{equation}
or
\begin{equation}
\Phi_\lambda = \frac{1}{1- \lambda \, (Q_B X) \, (\eta_0 J)} \,
\lambda \, X \,,
\label{solution}
\end{equation}
where
\begin{equation}
\frac{1}{1- \lambda \, (Q_B X) \, (\eta_0 J)}
\equiv 1 + \sum_{n=1}^\infty \,
[ \, \lambda \, ( Q_B X ) \, ( \eta_0 J ) \, ]^n \,.
\end{equation}

Let us now show that $\Phi_\lambda$ given by (\ref{solution})
satisfies the equation of motion (\ref{equation-of-motion}).
Since $Q_B X$ and $\eta_0 J$ are annihilated by $\eta_0$,
the state $\eta_0 \, \Phi_\lambda$ is given by
\begin{equation}
\eta_0 \, \Phi_\lambda
= \frac{1}{1- \lambda \, (Q_B X) \, (\eta_0 J)} \,
\lambda \, (\eta_0 X) \,.
\end{equation}
For the calculation of $Q_B \, \Phi_\lambda$,
we use $Q_B \, [ \, (Q_B X) \, (\eta_0 J) \, ]
= {}- Q_B X$ to find
\begin{equation}
Q_B \, \frac{1}{1- \lambda \, (Q_B X) \, (\eta_0 J)}
= {}- \frac{1}{1- \lambda \, (Q_B X) \, (\eta_0 J)} \,
\lambda \, (Q_B X) \,
\frac{1}{1- \lambda \, (Q_B X) \, (\eta_0 J)} \,.
\end{equation}
The state $Q_B \, \Phi_\lambda$ is given by
\begin{equation}
\begin{split}
Q_B \, \Phi_\lambda
= & {}- \frac{1}{1- \lambda \, (Q_B X) \, (\eta_0 J)} \,
\lambda \, (Q_B X) \,
\frac{1}{1- \lambda \, (Q_B X) \, (\eta_0 J)} \, \lambda \, X \\
& \quad {}+ \frac{1}{1- \lambda \, (Q_B X) \, (\eta_0 J)} \,
\lambda \, (Q_B X) \\
= & \, \frac{1}{1- \lambda \, (Q_B X) \, (\eta_0 J)} \,
\lambda \, (Q_B X) \, \Bigl[ \,
1 - \frac{1}{1- \lambda \, (Q_B X) \, (\eta_0 J)} \, \lambda \, X \,
\Bigr] \,.
\end{split}
\end{equation}
Note that
\begin{equation}
( \, Q_B \, \Phi_\lambda \, ) \, \frac{1}{1-\Phi_\lambda}
= \frac{1}{1- \lambda \, (Q_B X) \, (\eta_0 J)} \,
\lambda \, (Q_B X) \,.
\end{equation}
Finally, $Q_B \, \eta_0 \Phi_\lambda$ is given by
\begin{equation}
Q_B \, \eta_0 \, \Phi_\lambda
= {}- \frac{1}{1- \lambda \, (Q_B X) \, (\eta_0 J)} \,
\lambda \, (Q_B X) \,
\frac{1}{1- \lambda \, (Q_B X) \, (\eta_0 J)} \,
\lambda \, (\eta_0 X) \,.
\end{equation}
We have thus shown that $\Phi_\lambda$ given by (\ref{solution})
satisfies the equation of motion (\ref{equation-of-motion}).

An explicit expression of $\Phi^{(n)}$
in the CFT formulation is given by
\begin{equation}
\langle \, \varphi, \Phi^{(n)} \, \rangle
= \int_0^1 dt_1 \int_0^1 dt_2 \ldots \int_0^1 dt_{n-1} \,
\Bigl\langle \, f \circ \varphi (0) \,
\prod_{i=0}^{n-2}
\Bigl[ \, Q_B \cdot {\cal V} (1+ \ell_i) \, {\cal B} \,
\Bigr] \, {\cal V} (1+ \ell_{n-1}) \,
\Bigr\rangle_{{\cal W}_{1+\ell_{n-1}}} \,,
\end{equation}
where the BRST transformation of ${\cal V}$ is
\begin{equation}
Q_B \cdot {\cal V} (z)
= c V_1 (z) + \eta e^\phi V_{1/2} (z) \,.
\end{equation}
Note that ${\cal J}$ in $J$ has been replaced
by ${\cal B}$ in $\eta_0 J$ because of (\ref{eta_0-J}).
The term $\eta e^\phi V_{1/2} (1+\ell_i)$
in $Q_B \cdot {\cal V} (1+\ell_i)$ 
does not contribute when $i = 1,\, 2,\, \ldots \,, n-2$
because ${\cal B}^2 = 0 \,$.
By repeatedly using ${\cal B} \, c(z) \, {\cal B} = {\cal B}$,
we find
\begin{equation}
\begin{split}
\langle \, \varphi, \Phi^{(n)} \, \rangle
& = \int d^{n-1} t \,
\Bigl\langle \, f \circ \varphi (0) \,\,
c V_1 (1) \, {\cal B} \,
\prod_{i=1}^{n-2} \Bigl[ \, V_1 (1+ \ell_i) \, \Bigr] \,
c \, \xi e^{-\phi} V_{1/2} (1+ \ell_{n-1}) \,
\Bigr\rangle_{{\cal W}_{1+\ell_{n-1}}} \\
& \quad ~ + \int d^{n-1} t \,
\Bigl\langle \, f \circ \varphi (0) \,\,
\eta e^\phi V_{1/2} (1) \, {\cal B} \,
\prod_{i=1}^{n-2} \Bigl[ \, V_1 (1+ \ell_i) \, \Bigr] \,
c \, \xi e^{-\phi} V_{1/2} (1+ \ell_{n-1}) \,
\Bigr\rangle_{{\cal W}_{1+\ell_{n-1}}} \,,
\end{split}
\label{solution-CFT}
\end{equation}
where we have defined
\begin{equation}
\int d^{n-1} t
\equiv \int_0^1 dt_1 \int_0^1 dt_2 \ldots \int_0^1 dt_{n-1} \,.
\end{equation}

We can also construct a different solution
if we choose $\Phi^{(2)}$ to be $X \, (Q_B J) \, (\eta_0 X)$.
It is easy to show that $\overline{\Phi}_\lambda$ given by
\begin{equation}
\overline{\Phi}_\lambda = \lambda \, X \,
\frac{1}{1 - \lambda \, (Q_B J) \, (\eta_0 X)}
\label{conjugate-solution}
\end{equation}
satisfies the equation of motion (\ref{equation-of-motion}).
It is also straightforward to construct analytic solutions
based on star-algebra projectors other than the sliver state
using the method in \cite{Okawa:2006sn}.

\section{Regularity conditions}
\setcounter{equation}{0}

In the proof that the solution (\ref{solution}) satisfies
the equation of motion (\ref{equation-of-motion}),
we used the following relations:
\begin{equation}
\begin{split}
(Q_B X) \, (Q_B \eta_0 J) \, X & = (Q_B X) \, X \,, \\
(Q_B X) \, (Q_B \eta_0 J) \, (Q_B X) 
& = (Q_B X) \, (Q_B X) \,, \\
(Q_B X) \, (Q_B \eta_0 J) \, (\eta_0 X) 
& = (Q_B X) \, (\eta_0 X) \,.
\end{split}
\end{equation}
Let us study the conditions for these relations to hold.
Since
\begin{equation}
\begin{split}
\eta_0 \cdot {\cal V} (z)
= \eta_0 \cdot [ \, c \xi e^{-\phi} V_{1/2} (z) \, ]
& = -c e^{-\phi} V_{1/2} (z) \,, \\
Q_B \cdot {\cal V} (z)
= Q_B \cdot [ \, c \xi e^{-\phi} V_{1/2} (z) \, ]
& = c V_1 (z) + \eta e^\phi V_{1/2} (z) \,,
\end{split}
\end{equation}
and ${\cal V}$, $Q_B \cdot {\cal V}$, and $\eta_0 \cdot {\cal V}$
are all primary fields of dimension $0$,
the condition for (\ref{Q_B-eta_0-J}) gives
\begin{equation}
\begin{split}
\lim_{w \to z} \,
[ \, c V_1 (z) + \eta e^\phi V_{1/2} (z) \, ] \,
c \xi e^{-\phi} V_{1/2} (w) = 0 \,, \\
\lim_{w \to z} \,
[ \, c V_1 (z) + \eta e^\phi V_{1/2} (z) \, ] \,
[ \, c V_1 (w)
+ \eta e^\phi V_{1/2} (w) \, ] = 0 \,, \\
\lim_{w \to z} \,
[ \, c V_1 (z) + \eta e^\phi V_{1/2} (z) \, ] \,
c e^{-\phi} V_{1/2} (w) = 0 \,.
\end{split}
\end{equation}
These are satisfied if the operator products
$V_1 (z) \, V_{1/2} (w)$
and $V_1 (z) \, V_1 (w)$
are regular in the limit $w \to z$,
and $V_{1/2} (z) \, V_{1/2} (w)$ vanishes
in the limit $w \to z$.
The vertex operator $V_{1/2} (z)$ is Grassmann odd
so that the last condition is satisfied
if the operator product $V_{1/2} (z) \, V_{1/2} (w)$
is not singular.
To summarize, the equation of motion is satisfied
if the operator products
$V_1 (z) \, V_{1/2} (w)$,
$V_1 (z) \, V_1 (w)$,
and $V_{1/2} (z) \, V_{1/2} (w)$
are regular in the limit $w \to z$.

Let us next consider if the solution itself is finite
and if any intermediate steps in the proof are well defined.
The expressions can be divergent
when two or more operators collide,
but if the states
\begin{equation}
[ \, ( Q_B X ) \, ( \eta_0 J ) \, ]^{n-1} \, X \,, \qquad
[ \, ( Q_B X ) \, ( \eta_0 J ) \, ]^{n-1} \, ( Q_B X ) \,, \qquad
[ \, ( Q_B X ) \, ( \eta_0 J ) \, ]^{n-1} \, ( \eta_0 X )
\end{equation}
for any positive integer $n$ are finite,
the solution and any intermediate steps in the proof
are well defined.
An explicit expression
of $\Phi^{(n)} = [ \, ( Q_B X ) \, ( \eta_0 J ) \, ]^{n-1} \, X$
has been presented in~(\ref{solution-CFT}).
Expressions of
$[ \, ( Q_B X ) \, ( \eta_0 J ) \, ]^{n-1} \, ( Q_B X )$
and $[ \, ( Q_B X ) \, ( \eta_0 J ) \, ]^{n-1} \, ( \eta_0 X )$
can be obtained from (\ref{solution-CFT})
by replacing
$c \, \xi e^{-\phi} V_{1/2} (1+ \ell_{n-1})$
with
$c V_1 (1+ \ell_{n-1})
+ \eta e^\phi V_{1/2} (1+ \ell_{n-1})$
and with
$-c e^{-\phi} V_{1/2} (1+ \ell_{n-1}) \,$, respectively.
The $bc$ ghost sector is finite
because $c(z) \, {\cal B} \, c(w)$
is finite in the limit $w \to z$.
The superghost sector is also finite
because
$\eta e^\phi (1) \, \xi e^{-\phi} (1+ \ell_{n-1})$
and
$\eta e^\phi (1) \, \eta e^\phi (1+ \ell_{n-1})$
are finite in the limit $\ell_{n-1} \to 0$.
Therefore, all the expressions are well defined
if the contributions from the matter sector listed below
are finite:
\begin{equation}
\begin{split}
& \int_0^1 dt_1 \int_0^1 dt_2 \ldots \int_0^1 dt_{n-1} \,\,
\prod_{i=0}^{n-1}
\Bigl[ \, V_1 (1+ \ell_i) \, \Bigr] \,, \\
& \int_0^1 dt_1 \int_0^1 dt_2 \ldots \int_0^1 dt_{n-1} \,\,
V_{1/2} (1) \, \prod_{i=1}^{n-1}
\Bigl[ \, V_1 (1+ \ell_i) \, \Bigr] \,, \\
& \int_0^1 dt_1 \int_0^1 dt_2 \ldots \int_0^1 dt_{n-1} \,\,
\prod_{i=0}^{n-2}
\Bigl[ \, V_1 (1+ \ell_i) \, \Bigr] \,
V_{1/2} (1+ \ell_{n-1}) \,, \\
& \int_0^1 dt_1 \int_0^1 dt_2 \ldots \int_0^1 dt_{n-1} \,\,
V_{1/2} (1) \, \prod_{i=1}^{n-2}
\Bigl[ \, V_1 (1+ \ell_i) \, \Bigr] \,
V_{1/2} (1+ \ell_{n-1}) \,,
\end{split}
\end{equation}
where $\ell_i$ was defined in (\ref{ell-definition}).
To summarize,
if operator products of an arbitrary number of $V_1$'s
and at most two $V_{1/2}$'s are regular,
the solution (\ref{solution}) is well defined
and satisfies the equation of motion (\ref{equation-of-motion}).

\section{Conclusions and discussion}
\setcounter{equation}{0}

We have constructed analytic solutions for marginal deformations
in open superstring field theory
when operator products made of $V_1$'s and $V_{1/2}$'s are regular.
Our solutions are very simple
and remarkably similar to the solutions
in the bosonic case \cite{Schnabl:2007az, Kiermaier:2007ba}.
We expect that there will be further progress
of analytic methods in open superstring field theory.

It would be interesting to study the rolling tachyon
in open superstring field theory,
and we expect that marginal deformations
associated with the rolling tachyon solutions
satisfy the regularity conditions discussed
in the preceding section.
However, deformations we are interested in typically have
singular operator products of the marginal operator.
In the bosonic case, solutions to third order in $\lambda$
have been constructed when the operator product
of the marginal operator is singular~\cite{Kiermaier:2007ba}.
We hope that a procedure similar to the one
developed in the bosonic case will work in the superstring case,
and it is important to carry out the program
to all orders in the deformation parameter.

Our choice of $\Phi^{(2)}$ in (\ref{Phi^(2)})
was based on a technical reason,
and it is not clear if this gauge choice is physically suitable.
In particular, the solution $\Phi_\lambda$ in (\ref{solution})
does not satisfy the reality condition on the string field.
However, it is difficult for us to imagine
that there are two inequivalent solutions
generated by a single marginal operator
which coincide to linear order in $\lambda$,
and we expect that our solution is related to a real one
by a gauge transformation.
In fact, we can explicitly confirm this at $O(\lambda^2)$.
In order to see this,
it is useful to write the solution in the original definition
of the string field by inverting the field redefinition
(\ref{field-redefinition}):
\begin{equation}
\Phi_{old} = {}- \ln \, (\, 1 - \Phi_{new} \, )
= \sum_{n=1}^\infty \, \frac{1}{n} \, \Phi_{new}^n \,.
\end{equation}
We expand $\Phi_{old}$ in powers of $\lambda$ as
\begin{equation}
\Phi_{old} = \sum_{n=1}^\infty \, \lambda^n \, \Phi_{old}^{(n)} \,,
\end{equation}
and then $\Phi_{old}^{(2)}$ is given by
\begin{equation}
\Phi_{old}^{(2)} = \Phi_{new}^{(2)}
+ \frac{1}{2} \, ( \Phi_{new}^{(1)} )^2
= (Q_B X) \, (\eta_0 J) \, X
+ \frac{1}{2} \, X^2 \,.
\label{Phi_old^(2)}
\end{equation}
The string field $\Phi_{old}^{(2)}$ does not satisfy
the reality condition.\footnote{
A string field within our ansatz
satisfies the reality condition
when it is odd under the conjugation
given by replacing $X \to -X$
and by reversing the order of string products.
Signs from anticommuting Grassmann-odd string fields
have to be taken care of
in reversing the order of string products.
}
However, there is another solution
which satisfies the reality condition given by
\begin{equation}
\frac{1}{2} \, [ \,
(Q_B X) \, (\eta_0 J) \, X + X \, (\eta_0 J) \, (Q_B X) \, ] \,,
\label{another-Phi_old^(2)}
\end{equation}
and the difference between (\ref{Phi_old^(2)})
and (\ref{another-Phi_old^(2)}) is
\begin{equation}
(Q_B X) \, (\eta_0 J) \, X
+ \frac{1}{2} \, X^2
- \frac{1}{2} \, [ \,
(Q_B X) \, (\eta_0 J) \, X + X \, (\eta_0 J) \, (Q_B X) \, ]
= \frac{1}{2} \, Q_B \, [ \, X \, (\eta_0 J) \, X \, ]
\end{equation}
and can be eliminated by a gauge transformation.
The open superstring field theory formulated by Berkovits
can also be used to describe the $N=2$ string
by replacing $Q_B$ and $\eta_0$ with the generators
in the $N=2$ string \cite{Berkovits:1995ab},
but the reality condition on the string field
for the $N=2$ string does not seem to be satisfied
for $\Phi_\lambda$ in (\ref{solution}) either.\footnote{
Our understanding is that the conjugation
in \cite{Berkovits:1995ab} is given
by replacing $X \to X$, $J \to -J$, $Q_B \to \eta_0$,
and $\eta_0 \to Q_B$
and by reversing the order of string products,
and the string field should be even under the conjugation.
Again signs from anticommuting Grassmann-odd string fields
have to be taken care of
in reversing the order of string products.
The string field $\Phi_{new}$ in (\ref{field-redefinition})
is real when $\Phi_{old}$ is real
with respect to this reality condition,
while this is not the case for the reality condition
for the ordinary superstring discussed earlier.
}
The conjugation in \cite{Berkovits:1995ab}
seems to map $\Phi_\lambda$ in (\ref{solution})
to $\overline{\Phi}_\lambda$ in (\ref{conjugate-solution}).
We again expect that our solution is related
to a solution satisfying the reality condition
by a gauge transformation.
For example, ${}- (Q_B X) \, J \, (\eta_0 X)$,
which is another solution to the equation for $\Phi^{(2)}$,
seems to satisfy the reality condition,
and the difference between
${}- (Q_B X) \, J \, (\eta_0 X)$
and $\Phi^{(2)}$ in (\ref{Phi^(2)})
is $\eta_0 \, [ \, (Q_B X) \, J \, X \, ]$
and can be eliminated by a gauge transformation
generated by $\eta_0$.
We have also found that
$(Q_B X) \, (Q_B J) \, X \, (\eta_0 J) \, (\eta_0 X)$,
which seems to satisfy the reality condition,
solves the equation for $\Phi^{(3)}$
when $\Phi^{(2)}$ is ${}- (Q_B X) \, J \, (\eta_0 X)$,
but we have not been able to extend the solution
to all orders in $\lambda$.
We think that there is a good chance
that solutions satisfying the reality condition
for the ordinary superstring or for the $N=2$ string
can be found within our ansatz,
and it would be desirable to have their explicit expressions.
On the other hand,
we believe that the solution in (\ref{solution})
has an advantage because the actions of $Q_B$ and $\eta_0$
on (\ref{solution}) are very simple.

It has been expected that the moduli space of D-branes
are reproduced by the moduli space of solutions
to open string field theory,
and we think that our approach provides a concrete setup
to address this question.
We have seen a one-to-one correspondence
between the condition for exact marginality
in boundary CFT \cite{Recknagel:1998ih}
and the absence of obstruction
in solving the equation of motion
for string field theory
at $O(\lambda^2)$ in the bosonic case \cite{Kiermaier:2007ba}.
It would be important
to study the correspondence at higher orders
and in the superstring case,
and a better understanding
of the correspondence might help us complete
the program of constructing solutions
when the operator product of the marginal operator is singular.
We hope that further developments in this subject
will shed light on more conceptual issus in string theory
such as background independence
or the question why the condition
that the $\beta$ function vanishes in the world-sheet theory
gives the equation of motion in the spacetime theory.

\bigskip

\noindent
{\it Note added}

After the first version of this paper was submitted to arXiv,
we found analytic solutions satisfying the reality condition
\cite{Okawa:2007it}.
We also learned that T.~Erler independently constructed
analytic solutions satisfying the reality condition,
which were presented in the second version of \cite{Erler:2007rh}.

\newpage

\noindent
{\bf \large Acknowledgments}

\medskip

I would like to thank Volker Schomerus for helpful conversations.
I would also like to thank the Niels Bohr Institute in Copenhagen
for hospitality during part of this work.



\small

\begin{thebibliography}{99}

\bibitem{Witten:1985cc}
  E.~Witten,
  ``Noncommutative Geometry And String Field Theory,''
  Nucl.\ Phys.\  B {\bf 268}, 253 (1986).

\bibitem{Schnabl:2005gv}
  M.~Schnabl,
  ``Analytic solution for tachyon condensation
  in open string field theory,''
  Adv.\ Theor.\ Math.\ Phys.\  {\bf 10}, 433 (2006)
  [arXiv:hep-th/0511286].

\bibitem{Okawa:2006vm}
  Y.~Okawa,
  ``Comments on Schnabl's analytic solution
  for tachyon condensation in Witten's open string field theory,''
  JHEP {\bf 0604}, 055 (2006)
  [arXiv:hep-th/0603159].

\bibitem{Fuchs:2006hw}
  E.~Fuchs and M.~Kroyter,
  ``On the validity of the solution of string field theory,''
  JHEP {\bf 0605}, 006 (2006)
  [arXiv:hep-th/0603195].

\bibitem{Fuchs:2006an}
  E.~Fuchs and M.~Kroyter,
  ``Schnabl's ${\cal L}_0$ operator in the continuous basis,''
  JHEP {\bf 0610}, 067 (2006)
  [arXiv:hep-th/0605254].

\bibitem{Rastelli:2006ap}
  L.~Rastelli and B.~Zwiebach,
  ``Solving open string field theory with special projectors,''
  arXiv:hep-th/0606131.

\bibitem{Ellwood:2006ba}
  I.~Ellwood and M.~Schnabl,
  ``Proof of vanishing cohomology at the tachyon vacuum,''
  JHEP {\bf 0702}, 096 (2007)
  [arXiv:hep-th/0606142].

\bibitem{Fuji:2006me}
  H.~Fuji, S.~Nakayama and H.~Suzuki,
  ``Open string amplitudes in various gauges,''
  JHEP {\bf 0701}, 011 (2007)
  [arXiv:hep-th/0609047].

\bibitem{Fuchs:2006gs}
  E.~Fuchs and M.~Kroyter,
  ``Universal regularization for string field theory,''
  JHEP {\bf 0702}, 038 (2007)
  [arXiv:hep-th/0610298].

\bibitem{Okawa:2006sn}
  Y.~Okawa, L.~Rastelli and B.~Zwiebach,
  ``Analytic solutions for tachyon condensation
  with general projectors,''
  arXiv:hep-th/0611110.

\bibitem{Asano:2006hk}
  M.~Asano and M.~Kato,
  ``New covariant gauges in string field theory,''
  arXiv:hep-th/0611189.

\bibitem{Asano:2006hm}
  M.~Asano and M.~Kato,
  ``Level truncated tachyon potential in various gauges,''
  JHEP {\bf 0701}, 028 (2007)
  [arXiv:hep-th/0611190].

\bibitem{Erler:2006hw}
  T.~Erler,
  ``Split string formalism and the closed string vacuum,''
  arXiv:hep-th/0611200.

\bibitem{Imbimbo:2006tz}
  C.~Imbimbo,
  ``The spectrum of open string field theory
  at the stable tachyonic vacuum,''
  arXiv:hep-th/0611343.

\bibitem{Erler:2006ww}
  T.~Erler,
  ``Split string formalism and the closed string vacuum. II,''
  arXiv:hep-th/0612050.

\bibitem{Schnabl:2007az}
  M.~Schnabl,
  ``Comments on marginal deformations in open string field theory,''
  arXiv:hep-th/0701248.

\bibitem{Kiermaier:2007ba}
  M.~Kiermaier, Y.~Okawa, L.~Rastelli and B.~Zwiebach,
  ``Analytic solutions for marginal deformations
  in open string field theory,''
  arXiv:hep-th/0701249.

\bibitem{Sen:2000hx}
  A.~Sen and B.~Zwiebach,
  ``Large marginal deformations in string field theory,''
  JHEP {\bf 0010}, 009 (2000)
  [arXiv:hep-th/0007153].

\bibitem{Iqbal:2000qg}
  A.~Iqbal and A.~Naqvi,
  ``On marginal deformations in superstring field theory,''
  JHEP {\bf 0101}, 040 (2001)
  [arXiv:hep-th/0008127].

\bibitem{Takahashi:2001pp}
  T.~Takahashi and S.~Tanimoto,
  ``Wilson lines and classical solutions
  in cubic open string field  theory,''
  Prog.\ Theor.\ Phys.\  {\bf 106}, 863 (2001)
  [arXiv:hep-th/0107046].

\bibitem{Marino:2001ny}
  M.~Marino and R.~Schiappa,
  ``Towards vacuum superstring field theory: The supersliver,''
  J.\ Math.\ Phys.\  {\bf 44}, 156 (2003)
  [arXiv:hep-th/0112231].

\bibitem{Kluson:2002ex}
  J.~Kluson,
  ``Exact solutions of open bosonic string field theory,''
  JHEP {\bf 0204}, 043 (2002)
  [arXiv:hep-th/0202045].

\bibitem{Takahashi:2002ez}
  T.~Takahashi and S.~Tanimoto,
  ``Marginal and scalar solutions
  in cubic open string field theory,''
  JHEP {\bf 0203}, 033 (2002)
  [arXiv:hep-th/0202133].

\bibitem{Kluson:2002hr}
  J.~Kluson,
  ``Marginal deformations in the open bosonic string field theory
  for N D0-branes,''
  Class.\ Quant.\ Grav.\  {\bf 20}, 827 (2003)
  [arXiv:hep-th/0203089].

\bibitem{Kluson:2002av}
  J.~Kluson,
  ``Exact solutions in open bosonic string field theory
  and marginal deformation in CFT,''
  Int.\ J.\ Mod.\ Phys.\  A {\bf 19}, 4695 (2004)
  [arXiv:hep-th/0209255].

\bibitem{Kluson:2003xu}
  J.~Kluson,
  ``Exact solutions in SFT and marginal deformation in BCFT,''
  JHEP {\bf 0312}, 050 (2003)
  [arXiv:hep-th/0303199].

\bibitem{Coletti:2003ai}
  E.~Coletti, I.~Sigalov and W.~Taylor,
  ``Abelian and nonabelian vector field effective actions from
  string field theory,''
  JHEP {\bf 0309}, 050 (2003)
  [arXiv:hep-th/0306041].

\bibitem{Berkovits:2003ny}
  N.~Berkovits and M.~Schnabl,
  ``Yang-Mills action from open superstring field theory,''
  JHEP {\bf 0309}, 022 (2003)
  [arXiv:hep-th/0307019].

\bibitem{Sen:2004cq}
  A.~Sen,
  ``Energy momentum tensor and marginal deformations
  in open string field theory,''
  JHEP {\bf 0408}, 034 (2004)
  [arXiv:hep-th/0403200].

\bibitem{Katsumata:2004cc}
  F.~Katsumata, T.~Takahashi and S.~Zeze,
  ``Marginal deformations and closed string couplings
  in open string field theory,''
  JHEP {\bf 0411}, 050 (2004)
  [arXiv:hep-th/0409249].

\bibitem{Yang:2005iu}
  H.~Yang and B.~Zwiebach,
  ``Testing closed string field theory with marginal fields,''
  JHEP {\bf 0506}, 038 (2005)
  [arXiv:hep-th/0501142].

\bibitem{Kishimoto:2005bs}
  I.~Kishimoto and T.~Takahashi,
  ``Marginal deformations and classical solutions
  in open superstring field theory,''
  JHEP {\bf 0511}, 051 (2005)
  [arXiv:hep-th/0506240].

\bibitem{Taylor:2003gn}
  W.~Taylor and B.~Zwiebach,
  ``D-branes, tachyons, and string field theory,''
  arXiv:hep-th/0311017.

\bibitem{Sen:2004nf}
  A.~Sen,
  ``Tachyon dynamics in open string theory,''
  Int.\ J.\ Mod.\ Phys.\  A {\bf 20}, 5513 (2005)
  [arXiv:hep-th/0410103].

\bibitem{Rastelli:2005mz}
  L.~Rastelli,
  ``String field theory,''
  arXiv:hep-th/0509129.

\bibitem{Taylor:2006ye}
  W.~Taylor,
  ``String field theory,''
  arXiv:hep-th/0605202.

\bibitem{Rastelli:2000iu}
  L.~Rastelli and B.~Zwiebach,
  ``Tachyon potentials, star products and universality,''
  JHEP {\bf 0109}, 038 (2001)
  [arXiv:hep-th/0006240].

\bibitem{Zwiebach:1992ie}
  B.~Zwiebach,
  ``Closed string field theory: Quantum action
  and the B-V master equation,''
  Nucl.\ Phys.\  B {\bf 390}, 33 (1993)
  [arXiv:hep-th/9206084].

\bibitem{Saadi:1989tb}
  M.~Saadi and B.~Zwiebach,
  ``Closed String Field Theory from Polyhedra,''
  Annals Phys.\  {\bf 192}, 213 (1989).

\bibitem{Kugo:1989aa}
  T.~Kugo, H.~Kunitomo and K.~Suehiro,
  ``Nonpolynomial Closed String Field Theory,''
  Phys.\ Lett.\  B {\bf 226}, 48 (1989).

\bibitem{Kugo:1989tk}
  T.~Kugo and K.~Suehiro,
  ``Nonpolynomial Closed String Field Theory: Action
  And Its Gauge Invariance,''
  Nucl.\ Phys.\  B {\bf 337}, 434 (1990).

\bibitem{Kaku:1988zv}
  M.~Kaku,
  ``Geometric derivation of string field theory from first
  principles: Closed strings and modular invariance,''
  Phys.\ Rev.\  D {\bf 38}, 3052 (1988).

\bibitem{Kaku:1988zw}
  M.~Kaku and J.~D.~Lykken,
  ``Modular-invariant closed-string field theory,''
  Phys.\ Rev.\  D {\bf 38}, 3067 (1988).

\bibitem{Okawa:2004ii}
  Y.~Okawa and B.~Zwiebach,
  ``Heterotic string field theory,''
  JHEP {\bf 0407}, 042 (2004)
  [arXiv:hep-th/0406212].

\bibitem{Berkovits:2004xh}
  N.~Berkovits, Y.~Okawa and B.~Zwiebach,
  ``WZW-like action for heterotic string field theory,''
  JHEP {\bf 0411}, 038 (2004)
  [arXiv:hep-th/0409018].

\bibitem{Berkovits:1995ab}
  N.~Berkovits,
  ``SuperPoincare invariant superstring field theory,''
  Nucl.\ Phys.\  B {\bf 450}, 90 (1995)
  [Erratum-ibid.\  B {\bf 459}, 439 (1996)]
  [arXiv:hep-th/9503099].

\bibitem{Erler:2007rh}
  T.~Erler,
  ``Marginal Solutions for the Superstring,''
  arXiv:0704.0930.

\bibitem{Friedan:1985ge}
  D.~Friedan, E.~J.~Martinec and S.~H.~Shenker,
  ``Conformal Invariance, Supersymmetry And String Theory,''
  Nucl.\ Phys.\  B {\bf 271}, 93 (1986).

\bibitem{Polchinski:1998rr}
  J.~Polchinski,
  ``String theory. Vol. 2: Superstring theory and beyond,''
  {\it  Cambridge, UK: Univ. Pr. (1998) 531 p}

\bibitem{Berkovits:1999bs}
  N.~Berkovits and C.~T.~Echevarria,
  ``Four-point amplitude from open superstring field theory,''
  Phys.\ Lett.\  B {\bf 478}, 343 (2000)
  [arXiv:hep-th/9912120].

\bibitem{Recknagel:1998ih}
  A.~Recknagel and V.~Schomerus,
  ``Boundary deformation theory and moduli spaces of D-branes,''
  Nucl.\ Phys.\  B {\bf 545}, 233 (1999)
  [arXiv:hep-th/9811237].

\bibitem{Okawa:2007it}
  Y.~Okawa,
  ``Real analytic solutions for marginal deformations
  in open superstring field theory,''
  arXiv:0704.3612 [hep-th].

\end{thebibliography}
\end{document}